\documentclass[twocolumn,groupedaddress,
superscriptaddress,preprintnumbers,
amsmath,amssymb,
aps,prl
]{revtex4-1}
\usepackage{graphicx}
\usepackage{dcolumn}
\usepackage{bm}%
\usepackage[colorlinks=true,citecolor=blue]{hyperref}
\DeclareUnicodeCharacter{2032}{\ensuremath{'}}
\hypersetup{colorlinks=true,citecolor=blue,linkcolor=red,urlcolor=blue}

\usepackage{float}
\usepackage{amsmath}
 \usepackage[auto]{contour}

\usepackage{bm}
\newcommand{\beq}{\begin{equation}}
\newcommand{\eeq}{\end{equation}}
\newcommand{\beqnar}{\begin{eqnarray}}
\newcommand{\eeqnar}{\end{eqnarray}}
\newcommand{\bfig}{\begin{figure}}
\newcommand{\efig}{\end{figure}}

\usepackage{color}
\usepackage[dvipsnames]{xcolor} 
\usepackage{changes}

\newcommand{\dsigma}{\contour{black}{\color{white}$\sigma$}}

\begin{document}
\title{Nonlinear Optical Response in Pseudo-Hermitian Systems at Steady State}

\author{S. Sajad Dabiri}
\affiliation{Department of Physics, Zhejiang Normal University, Jinhua 321004, China}
\author{Reza Asgari}
\affiliation{Department of Physics, Zhejiang Normal University, Jinhua 321004, China}
\affiliation{School of Physics, Institute for Research in Fundamental Sciences (IPM), Tehran 19395-5531, Iran}
\date{\today}
\vspace{1cm}
\newbox\absbox
\begin{abstract}
We establish a steady-state theory for nonlinear optical conductivity in pseudo-Hermitian systems. We derive compact formulas for the first and second order conductivity tensors in both the velocity and length gauges and prove their exact equivalence through generalized sum rules and Berry connection identities by formulating the nonlinear response in terms of a biorthogonal density matrix. Utilizing the formalism on parity–time symmetric two-level systems reveals nonlinear phenomena that are not present in Hermitian systems, such as extra terms in the conductivity, corrections to the velocity operator, photocurrent, and resonance structures with higher-order poles at one-photon transitions. These features yield qualitatively distinct harmonic generation responses like real second-order conductivities and nonzero DC limits.  
These results provide new insights into nonlinear light–matter interactions in active media characterized by balanced gain and loss, with implications for non-Hermitian photonics, dissipative topological systems, and quantum devices designed with engineered dissipation.
\end{abstract}
\maketitle

\emph{\color{blue}Introduction}---
Non-Hermitian quantum systems with balanced gain and loss have attracted great interest for studying unconventional wave phenomena, exceptional points, and parity–time ($\mathcal{PT}$) symmetric physics \cite{moiseyev2011non,bergholtz2021exceptional,el2018non}. Despite rapid progress, a general framework for the nonlinear (NL) optical response of pseudo-Hermitian systems is still lacking. The linear regime has been recently developed for time-periodic systems \cite{seradjeh2020,dabiri3,PhysRevB.109.115431} and for non-Hermitian Hamiltonians \cite{bender1998real,mostafazadeh2002pseudo, ashida2020non, el2018non, PhysRevX.9.041015, xiao2025non}. In contrast, the NL response of Hermitian materials is well understood through perturbative density-matrix approaches \cite{taghi2017,watan2021,parker2019}. However, fundamental questions regarding gauge invariance, resonance structure, and steady-state behavior remain open in the non-Hermitian setting. Meden \textit{et al.} \cite{meden2023mathcal} established a fluctuation–dissipation relation \cite{PRXQuantum.3.030308} connecting non-Hermitian perturbations to Hermitian correlation functions, and Sticlet \textit{et al.} \cite{sticlet2022kubo} derived a Kubo formula for non-Hermitian Hamiltonians. Moreover, recent studies of multimode NL cavity systems combining Hermitian and non-Hermitian interactions have been exploited for controlling energy flow \cite{pontula2025non}. Nevertheless, the NL optical response in pseudo-Hermitian steady states remains unexplored.

Pseudo-Hermitian quantum mechanics provides a consistent framework for describing systems with balanced gain and loss, supporting unitary evolution and real spectra in the unbroken $\mathcal{PT}$ phase \cite{bender1998real,bender2007making, wang2023non,konotop2016nonlinear}. In this Letter, we develop a steady-state theory of NL optical conductivity (OC) in pseudo-Hermitian systems. Our main contributions are: (i) a general biorthogonal density-matrix formalism for the $n$th-order response in non-broken $\mathcal{PT}$ ($n\mathcal{PT}$) symmetric systems within the Schr\"{o}dinger picture; (ii) derivations of NL OC in both the velocity and length gauges, with their equivalence proved through generalized sum rules; and (iii) applications to paradigmatic models, including a $\mathcal{PT}$-symmetric qubit that exhibits unique structures in harmonic generation and gauge-invariant resonance phenomena.

Our results demonstrate that pseudo-Hermitian systems exhibit distinct NL features absent in Hermitian materials. These include additional contributions to the OC, modified sum rules, velocity-operator corrections, photocurrent generation, and unconventional resonance structures with higher-order poles. Such effects give rise to clear experimental signatures, including asymmetric harmonic generation spectra and finite DC nonlinear signals, which can be considered as sensitive probes of non-Hermitian physics. These findings open new avenues for NL photonics in $\mathcal{PT}$-symmetric waveguides \cite{guo2009observation,ruter2010observation}, exciton–polariton condensates, and dissipative topological materials \cite{bergholtz2021exceptional}.

 \emph{\color{blue}{Model and method}}---We begin with a time-independent, diagonalizable non-Hermitian Hamiltonian $H$. Diagonalizability ensures that the system is away from exceptional points, where two or more eigenvalues and their corresponding eigenstates coalesce. The right and left eigenvectors are defined as
$
{{H}}|{{R}_{\alpha }}\rangle ={{\epsilon }_{\alpha }}|{{R}_{\alpha }}\rangle$ and $,\langle {{L}_{\alpha }}|{{H}}={{\epsilon }_{\alpha }}\langle {{L}_{\alpha }}|
$, respectively.
The right eigenvectors are not orthonormal among themselves; instead, they satisfy a biorthogonality relation and a completeness condition, given by
$
\langle {{L}_{\alpha }}|{{R}_{\beta }}\rangle ={{\delta }_{\alpha \beta }}$ and $\mathop{\sum }_{\alpha }|{{R}_{\alpha }}\rangle \langle {{L}_{\alpha }}|=1 
$.
In general, if a non-Hermitian Hamiltonian with a discrete spectrum and a complete biorthogonal basis of eigenstates commutes with the parity–time operator, $[H,\mathcal{P}\mathcal{T}]=0$, it is pseudo-Hermitian. 
The eigenvalues are then either entirely real or occur in complex-conjugate pairs \cite{mostafazadeh2002pseudo,meden2023mathcal}. It has been shown that a $\mathcal{P}\mathcal{T}$-symmetric Hamiltonian possesses a fully real spectrum if and only if all of its right eigenvectors are also eigenvectors of $\mathcal{P}\mathcal{T}$ \cite{bender2005introduction,bender2007making}. We refer to this regime as $n\mathcal{P}\mathcal{T}$ symmetry.  Conversely, a Hamiltonian may commute with $\mathcal{P}\mathcal{T}$ while its eigenstates are not eigenstates of $\mathcal{P}\mathcal{T}$; in this case, the spectrum forms complex-conjugate pairs. 


The evolution of the density matrix is governed by the generalized von Neumann equation in  Schr\"odinger picture (working hereafter in natural units $e=\hbar=1$):
\begin{equation}
i{{\partial }_{t}}{{\rho }^{[0]}}(t)={{[H,{{\rho }^{[0]}}(t)]}_{\sim}}=H{{\rho }^{[0]}}(t)-{{\rho }^{[0]}}(t){{H}^{\dagger }}
\end{equation}
where ${\rho }^{[0]}$ denotes the density matrix in the absence of perturbation.  We neglect the quantum jumps term, which refers to a decoherence arising from the interaction between the physical system and the environment~\cite{xiao2025non,niu2023effect}. The density matrix for non-Hermitian systems can be defined in different ways in the framework of $\mathcal{PT}$-symmetric and biorthogonal formalism  \cite{bender2019pt,meden2023mathcal}. In the broken $\mathcal{PT}$ symmetric phase, because of complex eigenvalues, some components of the density matrix will grow or decay in time, avoiding the equilibrium situation. On the other hand, the most physical definition of density matrix can be given for  $n\mathcal{P}\mathcal{T}$ symmetric phase with a real spectrum as \cite{meden2023mathcal}
$
\rho _{n\mathcal{P}\mathcal{T}}^{[0]}(t)=\rho _{n\mathcal{P}\mathcal{T}}^{[0]}(0)=\mathop{\sum }_{\alpha }{{f}_{\alpha }}{|{{R}_{\alpha }}\rangle \langle {{R}_{\alpha }}|}/{\langle {{R}_{\alpha }}|{{R}_{\alpha }}\rangle }=\mathop{\sum }_{\alpha }{{\mathcal{F}}_{\alpha }}|{{R}_{\alpha }}\rangle \langle {{R}_{\alpha }}|,
\label{eqrho}
$
which is a diagonal stationary density matrix and does not evolve in time. $f_\alpha$ is a constant showing the occupation of states. In the following, we will obtain the response of $n\mathcal{PT}$- symmetric models described by the density matrix in the presence of an electromagnetic field.

 

The time evolution of the density matrix, after applying an electromagnetic field, can be derived from a generalized von Neumann equation \cite{meden2023mathcal} in the velocity and length gauge. 
Assuming $\lambda$ as a small constant and $\mathbf{A}(t)=\lambda \mathbf{V}(t)$, we may expand $\rho$ in powers of $\lambda$ as $\rho ={{\rho }^{[0]}}+\lambda {{\rho }^{[1]}}+{{\lambda }^{2}}{{\rho }^{[2]}}+...$. On the other hand, the Hamiltonian can be Taylor expanded in terms of powers of $\lambda$ as:
$
H(\mathbf{k}){{|}_{\mathbf{k}\to \mathbf{k}+\mathbf{A}(t)}}=h+{{h}^{i}}\lambda {{V}_{i}}(t)+\frac{{{h}^{ij}}{{\lambda }^{2}}{{V}_{i}}(t){{V}_{j}}(t)}{2!}+...
\label{totalh}
$
where summation over repeated indices should be carried out and $h\equiv H(\mathbf{k}),{{h}^{i}}\equiv {{\partial }_{{{k}_{i}}}}H(\mathbf{k}),{{h}^{ij}}\equiv {{\partial }_{{{k}_{i}}}}{{\partial }_{{{k}_{j}}}}H(\mathbf{k})$. We also assume that the derivatives of $h$ with respect to momentum are Hermitian, ensuring that the current is a well-defined observable. 

The velocity operator can be calculated by taking the derivative of the total Hamiltonian with respect to momentum \cite{parker2019,dabiri2025velocity} in both gauges \cite{note2}.
We can hence obtain the total optical current $
\langle\mathbf{J}\rangle=-{\text{Tr}({\bf v}\rho )}/{\text{Tr}(\rho )}$ in powers of $\lambda$; $\mathbf{J}={{\mathbf{J}}^{[0]}}+\lambda {{\mathbf{J}}^{[1]}}+{{\lambda }^{2}}{{\mathbf{J}}^{[2]}}+...$. 

\emph{\color{blue}{First-order optical conductivity (FOOC)}}-- Considering ${{V}_{z}}(t)={E{{e}^{-i{{\omega }_{1}}t}}}/{i{{\omega }_{1}}}$, we can determine the first-order density matrix \cite{note2} in each gauge. Therefore, the FOOC, as the proportionality coefficient of first-order current and electric field, can be expressed as 
\begin{equation}
\begin{aligned}
 {{\sigma }^{[1]}_{xz}}&=\frac{-\text{Tr}({{v}^{x[1]}}{{\rho }^{[0]}}+{{v}^{x[0]}}{{\rho }^{[1]}})+\text{Tr}({{v}^{x[0]}}{{\rho }^{[0]}})\text{Tr}({{\rho }^{[1]}})}{E{{e}^{-i\omega_1 t}}} \\ 
 & \equiv \sum\limits_{\mathbf{k}} \sigma _{\{100\}}^{[1]}+\sigma _{\{010\}}^{[1]}+\sigma _{\{001\}}^{[1]} 
\label{sig10}
\end{aligned}
\end{equation}
where the first-order density matrix in velocity gauge is $\rho _{{{L}_{\alpha }}{{L}_{\beta }}}^{[1]}=\frac{iE}{{{\omega }_{1}}}{\frac{{{e}^{-i{{\omega }_{1}}t}}}{{{\epsilon }_{\alpha \beta }}-{{\omega }_{1}}}({{\mathcal{F}}_{\beta }}h_{{{L}_{\alpha }}{{R}_{\beta }}}^{z}-{{\mathcal{F}}_{\alpha }}h_{{{R}_{\alpha }}{{L}_{\beta }}}^{z})}$, however, it becomes $ \frac{iE{{e}^{-i{{\omega }_{1}}t}}}{{{\omega }_{1}}}{{\delta }_{\alpha \beta }}{{\left( {{\mathcal{F}}_{\alpha }} \right)}_{;{{k}_{z}}}}  -E\frac{{{e}^{-i{{\omega }_{1}}t}}}{{{\epsilon }_{\alpha \beta }}-{{\omega }_{1}}}({{\mathcal{F}}_{\beta }}z_{{{L}_{\alpha }}{{R}_{\beta }}}^{e}-{{\mathcal{F}}_{\alpha }}z_{{{R}_{\alpha }}{{L}_{\beta }}}^{e})$ in the length gauge, where we define the generalized covariant derivative $  i{{\left( \mathcal{F}_\alpha \right)}_{;\mathbf{k}}}=i{{\partial }_{\mathbf{k}}}(\mathcal{F}_\alpha)+\mathcal{F}_\alpha({{\xi }_{{{L}_{\alpha }}{{R}_{\alpha }}}}-{{\xi }_{{{R}_{\alpha }}{{L}_{\alpha }}}})$, and generalized Berry connection ${{\xi }_{{{L}_{\alpha }}{{R}_{\beta }}}}=\langle {{L}_{\alpha }}(\mathbf{k})|i{{\partial }_{\mathbf{k}}}|{{R}_{\beta }}(\mathbf{k})\rangle $ with $z^e_{R_\alpha L_\beta}=\delta_{\alpha\beta}\xi^z_{R_\alpha L_\beta}$.
We define the matrix element ${{X}_{{{L}_{\alpha }}{{R}_{\beta }}}}=\langle {{L}_{\alpha }}|X|{{R}_{\beta }}\rangle$ and the various contributions to FOOC are given in \cite{note2}.
The sum over momentum $\mathbf{k}$ in Eq.~(\ref{sig10}) indicates the integral over Brillouin zone. 
The result closely parallels the Hermitian case \cite{parker2019}, except that for Hermitian Hamiltonians the term $\sigma_{{001}}^{[1]}$ in Eq.~(\ref{sig10}) vanishes. The extra term $\sigma _{\{001\}}^{[1]}$ can be interpreted as a correction to the velocity operator $h^x$ when combined with $\sigma _{\{010\}}^{[1]}$ \cite{note2}. This expression is consistent with previously reported results, although in Refs.~\cite{meden2023mathcal,sticlet2022kubo,pan2020non}, the authors did not consider the first term $\sigma _{\{100\}}^{[1]}$ of (\ref{sig10}) which stems from the variation of the velocity operator under the influence of perturbation, and therefore the first and last terms are missed in Ref.~\cite{tetling2022linear}. An important feature is that the FOOC of pseudo-Hermitian Hamiltonians with a real spectrum is time-independent: a probe field at frequency $\omega_1$ induces a response at the same frequency $\omega_1$, without any temporal growth or decay of the current.


\emph{\color{blue}{Second-order optical conductivity (SOOC)}}--
We can calculate the SOOC by first evaluating $\rho^{[2]}$ \cite{note2}, assuming ${{V}_{y}}(t)={E_2{{e}^{-i{{\omega }_{2}}t}}}/{i{{\omega }_{2}}}$. The SOOC is defined as
\begin{equation}
\begin{aligned}
  & {{\sigma }^{[2]}_{xyz}}=\frac{-\text{Tr}({{v}^{x[2]}}{{\rho }^{[0]}}+{{v}^{x[1]}}{{\rho }^{[1]}}+{{v}^{x[0]}}{{\rho }^{[2]}})}{E{{E}_{2}}{{e}^{-i({{\omega }_{1}}+{{\omega }_{2}})t}}} \\ 
 & +\frac{\text{Tr}({{v}^{x[0]}}{{\rho }^{[0]}})\text{Tr}({{\rho }^{[2]}})+\text{Tr}({{v}^{x[1]}}{{\rho }^{[0]}}+{{v}^{x[0]}}{{\rho }^{[1]}})\text{Tr}({{\rho }^{[1]}})}{E{{E}_{2}}{{e}^{-i({{\omega }_{1}}+{{\omega }_{2}})t}}} \\ 
 & \equiv\sum\limits_{\mathbf{k}}  \sigma _{\{200\}}^{[2]}+\sigma _{\{110\}}^{[2]}+\sigma _{\{020\}}^{[2]}+\sigma _{\{002\}}^{[2]}+\sigma _{\{101\}}^{[2]}+\sigma _{\{011\}}^{[2]}.
\label{sig20}
\end{aligned}
\end{equation}
where $\rho^{[2]}=\rho^{[2]vv} +\rho^{[2]v} $, $\rho _{{{L}_{\alpha }}{{L}_{\beta }}}^{[2]vv} =\frac{E{{E}_{2}}}{2{{\omega }_{1}}{{\omega }_{2}}}{\frac{{{e}^{i(-{{\omega }_{1}}-{{\omega }_{2}})t}}}{{{\epsilon }_{\alpha \beta }}-{{\omega }_{1}}-{{\omega }_{2}}}}\{ {{\mathcal{F}}_{\beta }}h_{{{L}_{\alpha }}{{R}_{\beta }}}^{yz}-{{\mathcal{F}}_{\alpha }}h_{{{R}_{\alpha }}{{L}_{\beta }}}^{yz} \}$ and $\rho _{{{L}_{\alpha }}{{L}_{\beta }}}^{[2]v}=-\frac{E{{E}_{2}}}{{{\omega }_{1}}{{\omega }_{2}}}\sum\limits_{\gamma }{\frac{{{e}^{i(-{{\omega }_{1}}-{{\omega }_{2}})t}}}{{{\epsilon }_{\alpha \beta }}-{{\omega }_{1}}-{{\omega }_{2}}}}\{ \frac{h_{{{L}_{\alpha }}{{R}_{\gamma }}}^{y}({{\mathcal{F}}_{\beta }}h_{{{L}_{\gamma }}{{R}_{\beta }}}^{z}-{{\mathcal{F}}_{\gamma }}h_{{{R}_{\gamma }}{{L}_{\beta }}}^{z})}{{{\epsilon }_{\gamma \beta }}-{{\omega }_{1}}}-\frac{h_{{{R}_{\gamma }}{{L}_{\beta }}}^{y}({{\mathcal{F}}_{\gamma }}h_{{{L}_{\alpha }}{{R}_{\gamma }}}^{z}-{{\mathcal{F}}_{\alpha }}h_{{{R}_{\alpha }}{{L}_{\gamma }}}^{z})}{{{\epsilon }_{\alpha \gamma }}-{{\omega }_{1}}} \} $. The form of the second-order density matrix in the length gauge is provided in \cite{note2}. 
The last three terms in Eq.~(\ref{sig20}) are specific to pseudo-Hermitian Hamiltonians and vanish identically in the Hermitian limit.
The SOOC is time-independent, and this property extends to response functions of arbitrary order. In pseudo-Hermitian systems, the SOOC contains additional contributions compared to the Hermitian case. These include a correction to the velocity operator, $h^x \rightarrow h^x - \langle h^x \rangle_0$, as implied by $\sigma _{\{002\}}^{[2]}$ in Eq.~(\ref{sig20}) as well as the one-photon resonance poles with doubled order as implied by $ \sigma _{\{101\}}^{[2]}$ and $\sigma _{\{011\}}^{[2]}$ in Eq.~(\ref{sig20}) \cite{note2}. This structure is consistent with the length gauge formulation. Notably, $\sigma^{[2]}$ expressions in Eq.~(\ref{sig20}) are not symmetrized, and the physical OC is obtained after symmetrizing them with respect to $(y,\omega_2)\leftrightarrow(z,\omega_1)$. 

\emph{\color{blue}{Applications}}--
We illustrate the utility of the perturbative formalism developed here by calculating the linear and NL optical responses of representative model systems. We begin with a $\mathcal{PT}$-symmetric single-qubit Hamiltonian 
\begin{equation}
\begin{aligned}
\mathcal{H}_\text{qubit}={\Delta}(\cos k {\sigma }_{x}-\sin k {\sigma }_{y})+i\gamma \sigma_z
\end{aligned}
\label{hqubit}
\end{equation}
where $\Delta=1$ is a real hopping/interaction term (Hermitian part), $\gamma$ is a real gain/loss parameter (non-Hermitian part), $\sigma_i,~i\in \{x,y,z,0\}$ is the Pauli matrix. The above Hamiltonian was experimentally realized by one trapped ion with tunable dissipation through microwave and lasers \cite{ding2021experimental}. Reference~\cite{fang2021experimental} experimentally investigated the coherence flow in such a qubit using optical setups. In this system, the parity operator is typically chosen as $\mathcal{P} = \sigma_x$, while the time-reversal operator $\mathcal{T} = \mathcal{K}$ corresponds to complex conjugation. The eigenvalues of Eq.~(\ref{hqubit}) are obtained as
  $ {{\epsilon }_{\pm }}=\pm \sqrt{1-{{\gamma }^{2}}}=\pm \zeta$.
As can be readily verified, biorthogonality holds except at the exceptional point ($\gamma=1$). For $\gamma < 1$ ($\gamma > 1$), the Hamiltonian (\ref{hqubit}) lies in the unbroken (broken) $\mathcal{PT}$ symmety phase. 

\begin{figure}
\includegraphics[width=\linewidth,trim={0 0 0.25cm 0}, clip]{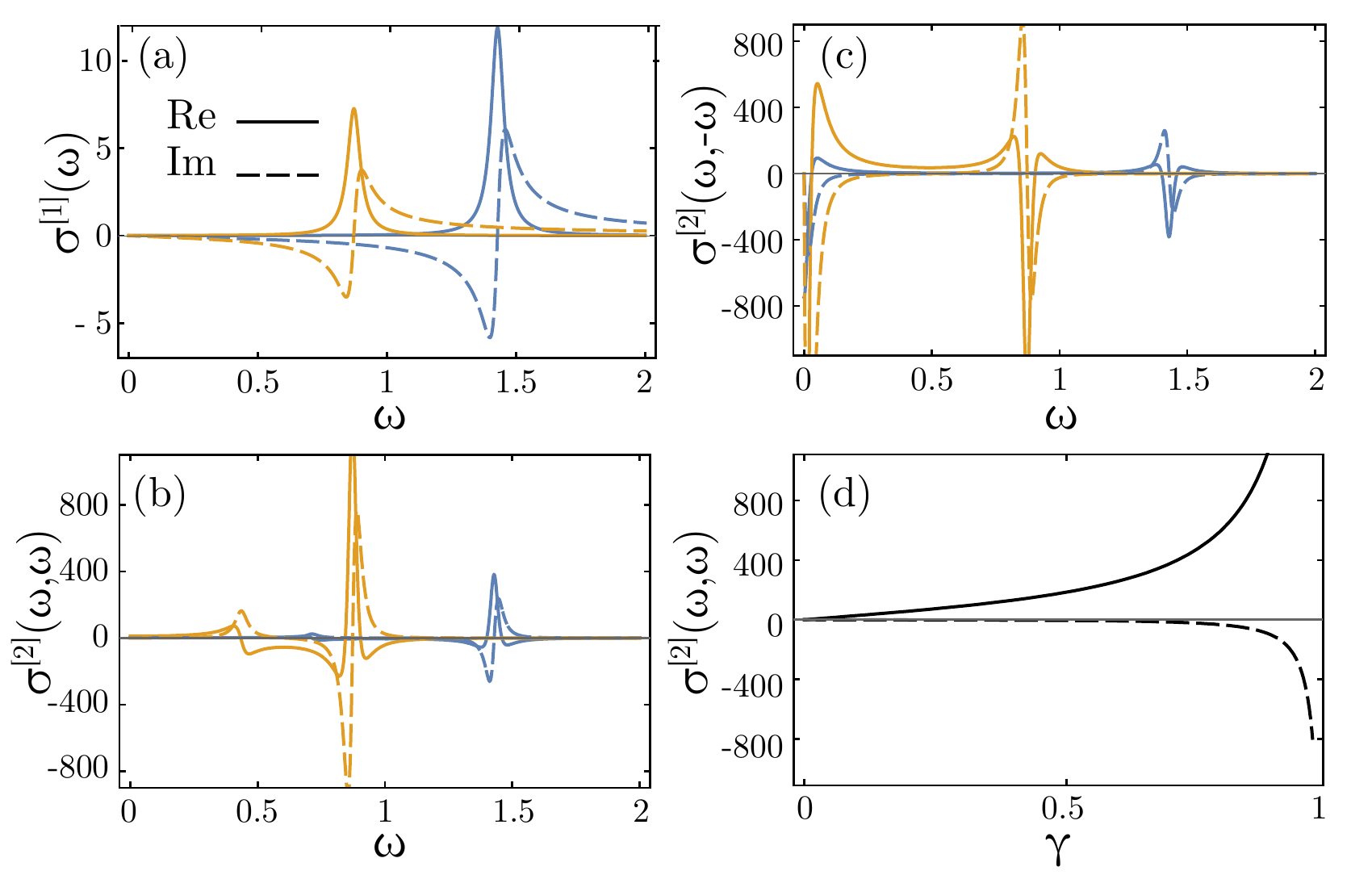} 
\caption{(Color online) (a) FOOC and (b),(c) SOOC of non-Hermitian qubit Hamiltonian (\ref{hqubit}) with parameters $\gamma=0.7$ (blue) and $\gamma=0.9$ (orange). Solid (dashed) lines depict the real (imaginary) parts. (d) SOOC as a function of $\gamma$ at one-photon resonance. According to (b),(d), as $\gamma$ increases toward the exceptional point ($\gamma = 1$), the SOOC resonance becomes sharper and more asymmetric, reflecting the approach to spectral degeneracy. The imaginary part undergoes the usual sign change at resonance, reminiscent of Hermitian absorption lines but modified by the non-Hermitian contributions to the velocity operator. Crucially, in (b) and (c), the one-photon resonance exhibits a second-order pole rather than the simple pole of Hermitian systems, producing sharp asymmetric line shapes and steep frequency dependence. The photocurrent in (c) remains finite, unlike the Hermitian qubit case, where it vanishes. This reflects the modified analytic structure of the conductivity tensor due to biorthogonal overlaps between left and right eigenstates. The second-harmonic generation is also real in the non-dissipative limit, in stark contrast with the purely imaginary Hermitian counterpart, implying a phase shift between fundamental and second-harmonic fields rather than absorption. This provides a clear experimental signature of pseudo-Hermiticity in NL optical measurements.  }
\label{qubitfig}
\end{figure}

The FOOC of qubit is obtained as ${{\sigma }^{[1]}_{n\mathcal{PT}}}=\frac{i\zeta \omega_1 }{{\omega_1^{2}}-4{{\zeta }^{2}}}. $ A peak in the OC appears at $\omega_1 = 2\zeta$, corresponding to the interband transition between the two levels. The linear response of the qubit is illustrated in Fig.~\ref{qubitfig}(a) for two different values of $\gamma$, where a small broadening to frequency $\omega \rightarrow \omega+i 0^+$ has been introduced to account for scattering. The real part shows a peak (corresponding to an absorption line) at a resonance frequency and remains positive, while the imaginary part changes sign at the resonance frequency, which is a typical response for Hermitian systems. The linear response calculated here agrees with previous results \cite{tetling2022linear} except that in Ref.~\cite{meden2023mathcal,tetling2022linear} the effect of changing the trace of the density matrix as a consequence of perturbation (Eq.~(\ref{sig10})) was neglected, which has a considerable effect. Let us show the Hermitian qubit Hamiltonian as follows $\mathcal{H}_\text{Her}={\Delta}(\cos k {\sigma }_{x}-\sin k {\sigma }_{y})-\theta \sigma_z,$
where $\theta\in \mathbb{R}$. The FOOC for Hermitian case is $\sigma^{[1]}_\text{Her}=\frac{\text{i}\omega }{\sqrt{1+{{\theta }^{2}}}\left( {{\omega }^{2}}-4(1+{{\theta }^{2}}) \right)}$ having a one-photon resonance and approaching to zero when $\omega \rightarrow 0$. 
To measure OC, the qubit can be coupled to a low-dissipation waveguide, and the transmission coefficients of electromagnetic waves can be measured \cite{macha2014implementation,jung2014multistability,fistul2022quantum}.

Next, we calculate the SOOC in velocity and length gauges by using Eqs.~(\ref{sig20})  and (\ref{hqubit}) for the second-harmonic generation (i.e. $\omega_1=\omega_2=\omega$) case as:
\begin{equation}
\begin{aligned}
{{\sigma }^{[2]}_{n\mathcal{PT}}}(\omega ,\omega )=\frac{\gamma \left( -8+8{{\gamma }^{4}}-2{{\gamma }^{2}}{{\omega }^{2}}-{{\omega }^{4}} \right)}{2\left( {{\omega }^{2}}-{{\zeta }^{2}} \right){{\left( {{\omega }^{2}}-4{{\zeta }^{2}} \right)}^{2}}}.
\end{aligned}
\label{sig2dirac}
\end{equation}
Equation (\ref{sig2dirac}) provides results in both the velocity and length gauges. There are resonances of one-photon and two-photons at $\omega=2\zeta$ and $\omega=\zeta$; moreover, there is no divergence at the zero frequency limit. In addition, Eq. (\ref{sig2dirac}) demonstrates that there is no SOOC at $\gamma=0$ because in that case the Hamiltonian has inversion symmetry $[H,\mathcal{P}]=[H,\sigma_x]=0$. 

For the sake of comparison, we calculate the SOOC of Hermitian qubit as $\sigma^{[2]}_\text{Her} (\omega,\omega )={3i\theta \omega }/\{{2\sqrt{1+{{\theta }^{2}}}\left( 1+{{\theta }^{2}}-{{\omega }^{2}} \right)\left( 4+4{{\theta }^{2}}-{{\omega }^{2}} \right)}\}$ which includes one-photon and two-photon resonances and approaches zero in zero frequency limit or zero staggered potential $\theta \rightarrow 0$ limit. Notably, both the one- and two-photon resonances exhibit poles of order $1$, leading to similar behavior of the OC in their vicinity.

The SOOC for the non-Hermitian qubit [Eq.~(\ref{sig2dirac})] is real, in contrast to the purely imaginary value found for its Hermitian counterpart. Within the slowly varying amplitude approximation and under the phase-matching condition \cite{boydbook}, the relation between the amplitudes of the second-harmonic beam ($E_{2\omega}$) and the fundamental beam ($E_\omega$) is given by $E_{2\omega} = \frac{-\sigma^{[2]}(\omega,\omega)}{2 n_{2\omega}c} E_\omega^2 z$, where $c$ is the speed of light, $z$ is the propagation length, and $n_{2\omega}$ is the refractive index at frequency $2\omega$. The real and imaginary parts of the optical conductivity determine the phase relationship between the second-harmonic and fundamental fields but are not directly related to absorption. The dissipated power for the second-harmonic beam is $P_{\text{diss}} = \frac{1}{2}\text{Re}(J_{2\omega}^* \cdot E_{2\omega}) = -\frac{1}{4c} \left| \sigma^{[2]}(\omega,\omega) E^2_\omega z \right| \text{Re}\left( \frac{1}{n_{2\omega}} \right)$. Typically, this dissipated power is negative, so the second-harmonic current amplifies the beam as it travels through the medium.

\begin{table*}[]
\caption{Key differences in the SOOC between Hermitian and pseudo-Hermitian qubits}
\begin{tabular}{llll}
\hline
$\sigma_{\text{Her}}(\omega,\omega)$ & $\sigma_{n\mathcal{PT}}(\omega,\omega)$ & $\sigma_{\text{Her}}(\omega,-\omega)$ & $\sigma_{n\mathcal{PT}}(\omega,-\omega)$ \\ \hline
Imaginary in non-dissipative case    & Real in non-dissipative case            & Zero in non-dissipative case          & Real in non-dissipative case             \\
Has only 1st-order poles             & Has 1st and 2nd-order poles             & No poles                              & Has only 2nd-order poles                 \\
Zero DC limit                        & Finite DC limit                       & Zero DC limit                         & Divergent DC limit                         
\end{tabular}
\label{tabcomp}
\end{table*}

 Fig.~\ref{qubitfig} (b) illustrates the SOOC of a single qubit obtained by using Eq.~(\ref{sig2dirac}). One-photon and two-photon resonances are observed as sharp changes in the OC. However, the structure of this response differs from typical responses in Hermitian systems: at one resonance, the derivative of the OC with respect to frequency reaches a minimum, while at the other resonance, it attains a maximum. This behavior arises because the two poles in Eq.~(\ref{sig2dirac}) have different orders i.e. the one-photon (two-photon) resonance poles order is equal to $2(1)$. 

We then proceed to calculate the SOOC associated with photocurrent generation (i.e., $\omega_1 = \omega$, $\omega_2 = -\omega$) in the length and velocity gauges, employing Eqs.~(\ref{sig20}) and (\ref{hqubit}). Remarkably, both gauges yield a unique result:
\begin{equation}
\begin{aligned}
\sigma _{n\mathcal{P}\mathcal{T}}^{[2]}(\omega ,-\omega )=\frac{4\gamma {{\zeta }^{2}}\left( 4{{\gamma }^{2}}+{{\omega }^{2}} \right)}{{{\omega }^{2}}{{\left( {{\omega }^{2}}-4{{\zeta }^{2}} \right)}^{2}}}.
\end{aligned}
\label{phpseudo}
\end{equation}
As seen from Eq.~(\ref{phpseudo}), second-order poles appear at the one-photon resonance frequency ($\omega = 2\zeta$) and in the DC limit ($\omega = 0$), shown in Fig.~\ref{qubitfig}(c). This behavior is in stark contrast to the Hermitian qubit case, where the DC photocurrent vanishes for all frequencies, i.e., $\sigma^{[2]}_{\text{Her}}(\omega,-\omega) = 0$. A summary of the key differences between the SOOC of Hermitian and pseudo-Hermitian qubits is provided in Table~\ref{tabcomp}.

Another notable point, as shown in Fig.~\ref{qubitfig}(d), is that the SOOC peak diverges when approaching the exceptional point at $\gamma = 1$, indicating high sensitivity. Such strong responses could be exploited in the design of high-performance experimental sensors \cite{chen2017exceptional}.

Finally, various effects that pseudo-Hermitian systems exhibit can be captured by different unique experimental signatures. To start, the resonance of a single photon appears as a second-order pole in $\sigma^{[2]}$, which can result in the second-order harmonic generation (SHG) spectrum displaying some exceptionally pronounced and asymmetrically shaped line features, unlike anything achievable in a Hermitian case. Moreover, the non-linear response loses the purely imaginary component, making the non-Hermitian system shift the phase relation between the fundamental and harmonic fields, unlike the purely imaginary OC of Hermitian systems. On the other hand, the DC limit of SOOC is non-zero for $n\mathcal{PT}$- symmetrix qubit as opposed to zero value for a Hermitian qubit. Moreover, the DC photocurrent in the pseudo-Hermitian qubit is nonzero, showing special resonances, while the Hermitian qubit has no finite photocurrents at a half-filled structure. 

\emph{\color{blue}{Conclusion}}--we have developed a unified and gauge-consistent treatment of the NL optical response of pseudo-Hermitian systems in the steady state. Using a biorthogonal and density matrix approach, we derived concise formulas for the NL response conductivity tensors in both the velocity and length gauges, and verified their exact equivalence through generalized sum rules. We identified NL phenomena that have no Hermitian counterpart, such as higher-order velocity operator corrections, extra resonance conductivity, and resonance phenomena of higher-order poles at the one-photon transition.

Realized non-Hermitian media, especially with genuinely non-vanishing DC limits, non-vanishing second-order conductivities, and with the NL response having a genuinely different analytic structure, are the first of many more non-Hermitian optical response media we may analyze. The theoretical implications of these results are definitive and, most importantly, unmistakable are the experimental resolutions, like sharply asymmetric lines and abrupt shifts in the optical response and in the harmonic generation phase that may be observed in $\mathcal{P}\mathcal{T}$ symmetric optical devices, engineered dissipative systems, and exciton-polariton condensates. This work opens the exploration of the NL optical response in active and topological non-Hermitian materials.

\nocite{apsrev41Control}
\bibliographystyle{apsrev4-1}
\bibliography{ref}

\newpage
\appendix

\section{Supplementary Material\\ Nonlinear Optical Response in Pseudo-Hermitian Systems at Steady State
}\label{detaild}
In this Supplementary Material, we explicitly derive the density matrices and provide proofs for the OC formulas presented in the main text.

\subsection{velocity gauge}\label {A0}
The initial density matrix is defined as 
\begin{equation}
\begin{aligned}\rho _{n\mathcal{P}\mathcal{T}}^{[0]}(t)=\rho _{n\mathcal{P}\mathcal{T}}^{[0]}(0)=\mathop{\sum }_{\alpha }{{f}_{\alpha }}\frac{{|{{R}_{\alpha }}\rangle \langle {{R}_{\alpha }}|}}{{\langle {{R}_{\alpha }}|{{R}_{\alpha }}\rangle }}=\mathop{\sum }_{\alpha }{{\mathcal{F}}_{\alpha }}|{{R}_{\alpha }}\rangle \langle {{R}_{\alpha }},
\end{aligned}
\label{eqrho}
\end{equation}
which is diagonal in the left eigenbasis. The occupation of states in thermal equilibrium can be taken as the Boltzmann factors $f_\alpha \propto e^{-E_\alpha/k_BT}$ with $k_B(T)$ being the Boltzmann constant (absolute temperature) with a proper normalization $\sum_\alpha f_\alpha=1$. By substituting the expansion of Hamiltonian and density matrix in generalized von Neuman equation $i{{\partial }_{t}}\rho (\mathbf{k},t)={{[H(\mathbf{k}){{|}_{\mathbf{k}\to \mathbf{k}+\mathbf{A}(t)}},\rho (\mathbf{k},t)]}_{\sim}}$ and suppressing the $\mathbf{k}$ index for brevity, we find 
\begin{equation}
\begin{aligned}
 i{{\partial }_{t}}{{\rho }^{[0]}}=  &[h,{{\rho }^{[0]}}]_{\sim}, \\ 
 i{{\partial }_{t}}{{\rho }^{[1]}}=  &[h,{{\rho }^{[1]}}]_{\sim}+[{{h}^{i}}{{V}_{i}}(t),{{\rho }^{[0]}}]_{\sim}, \\ 
 i{{\partial }_{t}}{{\rho }^{[2]}}=  &[h,{{\rho }^{[2]}}]_{\sim}+[{{h}^{i}}{{V}_{i}}(t),{{\rho }^{[1]}}]_{\sim}+[\frac{{{h}^{ij}}}{2!}{{V}_{i}}(t){{V}_{j}}(t),{{\rho }^{[0]}}]_{\sim}, \\ 
  i{{\partial }_{t}}{{\rho }^{[3]}}=  &[h,{{\rho }^{[3]}}]_{\sim}+[{{h}^{i}}{{V}_{i}}(t),{{\rho }^{[2]}}]_{\sim}+[\frac{{{h}^{ij}}}{2!}{{V}_{i}}(t){{V}_{j}}(t),{{\rho }^{[1]}}]_{\sim} \\ 
 & +[\frac{{{h}^{ijl}}}{3!}{{V}_{i}}(t){{V}_{j}}(t){{V}_{l}}(t),{{\rho }^{[0]}}]_{\sim}.
\end{aligned}
\label{3rho}
\end{equation}
Using the assumption of the Hermiticity of the velocity operator and the equilibrium initial density matrix (\ref{eqrho}), Eqs (\ref{3rho}) can be written more compactly:
\begin{equation}
\begin{aligned}
& i{{\partial }_{t}}{{\rho }^{[0]}}=0 \\ 
 & i{{\partial }_{t}}\rho _{D}^{[1]}={{[h_{D}^{i}{{V}_{i}}(t),{{\rho }^{[0]}}]}_{\sim}} \\ 
 & i{{\partial }_{t}}\rho _{D}^{[2]}={{[h_{D}^{i}{{V}_{i}}(t),\rho _{D}^{[1]}]}_{\sim}}+{{[\frac{h_{D}^{ij}}{2!}{{V}_{i}}(t){{V}_{j}}(t),{{\rho }^{[0]}}]}_{\sim}} \\ 
 & i{{\partial }_{t}}\rho _{D}^{[3]}={{[h_{D}^{i}{{V}_{i}}(t),\rho _{D}^{[2]}]}_{\sim}}+{{[\frac{h_{D}^{ij}}{2!}{{V}_{i}}(t){{V}_{j}}(t),\rho _{D}^{[1]}]}_{\sim}} \\ 
 & +{{[\frac{h_{D}^{ijl}}{3!}{{V}_{i}}(t){{V}_{j}}(t){{V}_{l}}(t),{{\rho }^{[0]}}]}_{\sim}},
\end{aligned}
\label{rhodha}
\end{equation}
where we have defined the Dirac representation of density matrix $\rho _{D}^{\,}={{e}^{iht}}\rho {{e}^{-i{{h}^{\dagger }}t}}$ and operator ${{A}_{D}}={{e}^{iht}}A{{e}^{-iht}}$, also the generalized commutator is defined as ${{[h_{D}^{z},\rho ]}_{\sim}}=h_{D}^{z}\rho -\rho h_{D}^{z\dagger }={{e}^{iht}}{{h}^{z}}{{e}^{-iht}}\rho -\rho {{e}^{i{{h}^{\dagger }}t}}{{h}^{z}}{{e}^{-i{{h}^{\dagger }}t}}$. Using Eq.~(\ref{eqrho}), it is clear that $\rho _{D}^{[0]}=\rho _{\,}^{[0]}$.

\subsubsection{First-order optical conductivity (FOOC)}\label{A1}
The left matrix elements of the density operators will be derived from Eq.~(\ref{rhodha}) in a rather straightforward manner. Assuming ${{V}_{z}}(t)=\frac{E{{e}^{-i{{\omega }_{1}}t}}}{i{{\omega }_{1}}}$ in the second equation of (\ref{rhodha}) and integrating gives
\begin{equation}
\begin{aligned}
&\rho _{{{L}_{\alpha }}{{L}_{\beta }}}^{[1]}=-\frac{i{{e}^{-i{{\epsilon }_{\alpha \beta }}t}}E}{i{{\omega }_{1}}}\sum\limits_{\gamma }{\int_{-\infty }^{t}{{{e}^{i({{\epsilon }_{\alpha \beta }}-{{\omega }_{1}})t'}}}} \\ 
 &~~~~~~~~~~~~~~~~~~~~~~~~~~\times \left( h_{{{L}_{\alpha }}{{R}_{\gamma }}}^{z}\rho _{{{L}_{\gamma }}{{L}_{\beta }}}^{[0]}-\rho _{{{L}_{\alpha }}{{L}_{\gamma }}}^{[0]}h_{{{R}_{\gamma }}{{L}_{\beta }}}^{z\dagger } \right)d{{t}^{\prime }} \\ 
  &=-\frac{{{e}^{-i{{\epsilon }_{\alpha \beta }}t}}E}{{{\omega }_{1}}}\int_{-\infty }^{t}{{{e}^{i({{\epsilon }_{\alpha \beta }}-{{\omega }_{1}})t'}}({{\mathcal{F}}_{\beta }}h_{{{L}_{\alpha }}{{R}_{\beta }}}^{z}-{{\mathcal{F}}_{\alpha }}h_{{{R}_{\alpha }}{{L}_{\beta }}}^{z})d{{t}^{\prime }}}. \\ 
\end{aligned}
\label{}
\end{equation}
Therefore, the matrix elements of the first-order density matrix is
\begin{equation}
\begin{aligned}
\rho _{{{L}_{\alpha }}{{L}_{\beta }}}^{[1]}=\frac{iE}{{{\omega }_{1}}}{\frac{{{e}^{-i{{\omega }_{1}}t}}}{{{\epsilon }_{\alpha \beta }}-{{\omega }_{1}}}({{\mathcal{F}}_{\beta }}h_{{{L}_{\alpha }}{{R}_{\beta }}}^{z}-{{\mathcal{F}}_{\alpha }}h_{{{R}_{\alpha }}{{L}_{\beta }}}^{z})}.
\end{aligned}
\label{rho1eq0}
\end{equation}
Given the first-order density matrix, it is an easy task to derive the explicit formula for FOOC elements of Eq.~(\ref{sig10}) as
\begin{equation}
\begin{aligned}
  & \sigma _{\{100\}}^{[1]}=\sum\limits_{\alpha}{\,}\frac{i}{{{\omega }_{1}}}{{\mathcal{F}}_{\alpha }}h_{{{R}_{\alpha }}{{R}_{\alpha }}}^{xz}, \\ 
 & \sigma _{\{010\}}^{[1]}=-\frac{i}{{{\omega }_{1}}}\sum\limits_{\alpha \beta}{\frac{h_{{{R}_{\beta }}{{R}_{\alpha }}}^{x}({{\mathcal{F}}_{\beta }}h_{{{L}_{\alpha }}{{R}_{\beta }}}^{z}-{{\mathcal{F}}_{\alpha }}h_{{{R}_{\alpha }}{{L}_{\beta }}}^{z})}{{{\epsilon }_{\alpha \beta }}-{{\omega }_{1}}}}, \\ 
 & \sigma _{\{001\}}^{[1]}=\sum\limits_{\alpha \beta}{\frac{i{{\langle {{h}^{x}}\rangle }_{0}}}{{{\omega }_{1}}}\frac{{{\mathbf{1}}_{{{R}_{\beta }}{{R}_{\alpha }}}}({{\mathcal{F}}_{\beta }}h_{{{L}_{\alpha }}{{R}_{\beta }}}^{z}-{{\mathcal{F}}_{\alpha }}h_{{{R}_{\alpha }}{{L}_{\beta }}}^{z})}{{{\epsilon }_{\alpha \beta }}-{{\omega }_{1}}}}, \\ 
 & {{\langle {{h}^{x}}\rangle }_{0}}=\sum\limits_{\gamma }{{{\mathcal{F}}_{\gamma }}h_{{{R}_{\gamma }}{{R}_{\gamma }}}^{x}.} \\ 
\label{sig1n}
\end{aligned}
\end{equation}

\subsubsection{Second-order optical conductivity (SOOC)}\label{A2}
We evaluate the second-order density matrix by using the third equation of (\ref{rhodha}) and Eq.~(\ref{eqrho}).  By defining ${{V}_{y}}(t)={{{E}_{2}}{{e}^{-i{{\omega }_{2}}t}}}/{i{{\omega }_{2}}}$, and $\rho^{[2]}=\rho^{[2]v}+\rho^{[2]vv}$, we have
\begin{widetext}

\begin{eqnarray}
&& \rho _{{{L}_{\alpha }}{{L}_{\beta }}}^{[2]v}=-\frac{i{{e}^{-i{{\epsilon }_{\alpha \beta }}t}}{{E}_{2}}}{i{{\omega }_{2}}}\sum\limits_{\gamma }{\int_{-\infty }^{t}{{{e}^{i({{\epsilon }_{\alpha \beta }}-{{\omega }_{2}})t'}}}}\times \left( h_{{{L}_{\alpha }}{{R}_{\gamma }}}^{y}\rho _{{{L}_{\gamma }}{{L}_{\beta }}}^{[1]}-\rho _{{{L}_{\alpha }}{{L}_{\gamma }}}^{[1]}h_{{{R}_{\gamma }}{{L}_{\beta }}}^{y} \right)d{{t}^{\prime }}\nonumber \\ 
&& \rho _{{{L}_{\alpha }}{{L}_{\beta }}}^{[2]v}=-\frac{E{{E}_{2}}}{{{\omega }_{1}}{{\omega }_{2}}}\sum\limits_{\gamma }{\frac{{{e}^{i(-{{\omega }_{1}}-{{\omega }_{2}})t}}}{{{\epsilon }_{\alpha \beta }}-{{\omega }_{1}}-{{\omega }_{2}}}}\left[ \frac{h_{{{L}_{\alpha }}{{R}_{\gamma }}}^{y}({{\mathcal{F}}_{\beta }}h_{{{L}_{\gamma }}{{R}_{\beta }}}^{z}-{{\mathcal{F}}_{\gamma }}h_{{{R}_{\gamma }}{{L}_{\beta }}}^{z})}{{{\epsilon }_{\gamma \beta }}-{{\omega }_{1}}}-\frac{h_{{{R}_{\gamma }}{{L}_{\beta }}}^{y}({{\mathcal{F}}_{\gamma }}h_{{{L}_{\alpha }}{{R}_{\gamma }}}^{z}-{{\mathcal{F}}_{\alpha }}h_{{{R}_{\alpha }}{{L}_{\gamma }}}^{z})}{{{\epsilon }_{\alpha \gamma }}-{{\omega }_{1}}} \right] 
\label{rho2v} \\
&& \rho _{{{L}_{\alpha }}{{L}_{\beta }}}^{[2]vv}=-\frac{i{{e}^{-i{{\epsilon }_{\alpha \beta }}t}}E{{E}_{2}}}{2i{{\omega }_{2}}i{{\omega }_{1}}}\sum\limits_{\gamma }{\int_{-\infty }^{t}{{{e}^{i({{\epsilon }_{\alpha \beta }}-{{\omega }_{1}}-{{\omega }_{2}})t'}}}}\times \left( h_{{{L}_{\alpha }}{{R}_{\gamma }}}^{yz}\rho _{{{L}_{\gamma }}{{L}_{\beta }}}^{[0]}-\rho _{{{L}_{\alpha }}{{L}_{\gamma }}}^{[0]}h_{{{R}_{\gamma }}{{L}_{\beta }}}^{yz} \right)d{{t}^{\prime }} \nonumber\\ 
&& \rho _{{{L}_{\alpha }}{{L}_{\beta }}}^{[2]vv}=\frac{E{{E}_{2}}}{2{{\omega }_{1}}{{\omega }_{2}}}{\frac{{{e}^{i(-{{\omega }_{1}}-{{\omega }_{2}})t}}}{{{\epsilon }_{\alpha \beta }}-{{\omega }_{1}}-{{\omega }_{2}}}}\left[ {{\mathcal{F}}_{\beta }}h_{{{L}_{\alpha }}{{R}_{\beta }}}^{yz}-{{\mathcal{F}}_{\alpha }}h_{{{R}_{\alpha }}{{L}_{\beta }}}^{yz} \right]. 
\label{rho2vv}
\end{eqnarray}
\end{widetext}
Using the biorthogonal basis and the first and second order density matrices in Eq.~(\ref{rho1eq0}),(\ref{rho2v}) and (\ref{rho2vv}), it is an easy step to find the explicit formulas for elemets of Eq.~(\ref{sig20}) SOOC as
 \begin{eqnarray}
\begin{aligned}
   & \sigma _{\{200\}}^{[2]}=\sum\limits_{\alpha }{\,}\frac{h_{{{R}_{\alpha }}{{R}_{\alpha }}}^{xyz}{{\mathcal{F}}_{\alpha }}}{2{{\omega }_{1}}{{\omega }_{2}}}, \\ 
 & \sigma _{\{110\}}^{[2]}=-\sum\limits_{\alpha \beta }{\,}\frac{h_{{{R}_{\beta }}{{R}_{\alpha }}}^{xy}}{{{\epsilon }_{\alpha \beta }}-{{\omega }_{1}}}\frac{({{\mathcal{F}}_{\beta }}h_{{{L}_{\alpha }}{{R}_{\beta }}}^{z}-{{\mathcal{F}}_{\alpha }}h_{{{R}_{\alpha }}{{L}_{\beta }}}^{z})}{{{\omega }_{1}}{{\omega }_{2}}}, \\ 
 & \sigma _{\{020\}}^{[2]}=-\frac{\sum\limits_{\alpha \beta }{h_{{{R}_{\beta }}{{R}_{\alpha }}}^{x}\rho_{{{L}_{\alpha }}{{L}_{\beta }}}^{[2]}}}{E{{E}_{2}}{{e}^{-i({{\omega }_{1}}+{{\omega }_{2}})t}}}\text{ }, \\ 
 & \sigma _{\{002\}}^{[2]}=\frac{\sum\limits_{\alpha \beta }{{{\langle {{h}^{x}}\rangle }_{0}}\mathbf{1}_{{{R}_{\beta }}{{R}_{\alpha }}} \rho _{{{L}_{\alpha }}{{L}_{\beta }}}^{[2]}}}{E{{E}_{2}}{{e}^{-i({{\omega }_{1}}+{{\omega }_{2}})t}}}\text{ }, \\ 
 & \sigma _{\{101\}}^{[2]}=-\sigma _{\{100\}}^{[1]}{{\text{ }\!\!|\!\!\text{ }}_{\text{(}z,{{\omega }_{1}})\to \text{(}y,{{\omega }_{2}})}}\frac{\text{Tr}({{\rho }^{[1]}})}{E{{e}^{-i{{\omega }_{1}}t}}}, \\ 
 & \sigma _{\{011\}}^{[2]}=-\sigma _{\{010\}}^{[1]}{{\text{ }\!\!|\!\!\text{ }}_{\text{(}z,{{\omega }_{1}})\to \text{(}y,{{\omega }_{2}})}}\frac{\text{Tr}({{\rho }^{[1]}})}{E{{e}^{-i{{\omega }_{1}}t}}}.
\label{sigmaha}
  \end{aligned}
\end{eqnarray}


\subsection{length gauge}
Now, let us consider the length gauge and find the optical conductivity formulas. The generalized von Neumann equation governing the evolution of the density matrix can be written as
\begin{equation}
i{{\partial }_{t}}\varrho (\mathbf{k},t)={{[H(\mathbf{k})+\lambda \mathcal{V}(t),\varrho (\mathbf{k},t)]}_{\sim}},
\label{eqneul}
\end{equation}
where $\lambda \mathcal{V}(t)=\mathbf{r}.\mathbf{E}(t)$, so the added perturbation is assumed to be Hermitian. After expanding the density matrix in terms of a small parameter $\lambda$  and substituting into Eq.~(\ref{eqneul}). The recursive relation between density matrices of different orders can be obtained through straightforward calculations and is given by:  
\begin{equation}
\varrho _{{{L}_{\alpha }}{{L}_{\beta }}}^{[n]}=-i{{e}^{-i{{\epsilon }_{\alpha \beta }}t}}\int_{-\infty }^{t}{{{e}^{i{{\epsilon }_{\alpha \beta }}t'}}\langle {{L}_{\alpha }}|[\mathcal{V}({{t}^{\prime }}),{{\varrho }^{[n-1]}}]|{{L}_{\beta }}\rangle d{{t}^{\prime }}},
\label{rhon}
\end{equation}
where $n=0,1,...$ and $\varrho^{[-1]}=0$. In order to evaluate the above equation, we need to find out the commutator of the position operator and the density matrix. Using the fact that the position operator acts as the derivative with respect to momentum ($\mathbf{r}\rightarrow i\partial_\mathbf{k}$), we can calculate the commutator by splitting the position operator into intraband and interband parts $\mathbf{r}\equiv\mathbf{r}^i+\mathbf{r}^e$ \cite{sipe1995}. We generalize this concept to the case of a pseudo-Hermitian system with a real spectrum by writing 
\begin{equation}
\begin{aligned}
  & {{[\mathbf{r},\varrho ]}_{{{L}_{\alpha }}{{L}_{\beta }}}}={{[{{\mathbf{r}}^{i}},\varrho ]}_{{{L}_{\alpha }}{{L}_{\beta }}}}+{{[{{\mathbf{r}}^{e}},\varrho ]}_{{{L}_{\alpha }}{{L}_{\beta }}}}, \\ 
 & {{[{{\mathbf{r}}^{i}},\varrho ]}_{{{L}_{\alpha }}{{L}_{\beta }}}}=i{{\partial }_{\mathbf{k}}}({{\varrho }_{{{L}_{\alpha }}{{L}_{\beta }}}})+{{\varrho }_{{{L}_{\alpha }}{{L}_{\beta }}}}({{\xi }_{{{L}_{\alpha }}{{R}_{\alpha }}}}-{{\xi }_{{{R}_{\beta }}{{L}_{\beta }}}}), \\ 
 & \equiv i{{\left( {{\varrho }_{{{L}_{\alpha }}{{L}_{\beta }}}} \right)}_{;\mathbf{k}}}, \\ 
 & {{[{{\mathbf{r}}^{e}},\varrho ]}_{{{L}_{\alpha }}{{L}_{\beta }}}}=\sum\limits_{\gamma }{\mathbf{r}_{{{L}_{\alpha }}{{R}_{\gamma }}}^{e}{{\varrho }_{{{L}_{\gamma }}{{L}_{\beta }}}}-{{\varrho }_{{{L}_{\alpha }}{{L}_{\gamma }}}}\mathbf{r}_{{{R}_{\gamma }}{{L}_{\alpha }}}^{e}.} 
\label{rire}
\end{aligned}
\end{equation}
where we have defined the generalized Berry connection ${{\xi }_{{{L}_{\alpha }}{{R}_{\beta }}}}=\langle {{L}_{\alpha }}(\mathbf{k})|i{{\partial }_{\mathbf{k}}}|{{R}_{\beta }}(\mathbf{k})\rangle $ and the generalized covariant derivative ${{\left( {{\varrho }_{{{L}_{\alpha }}{{L}_{\beta }}}} \right)}_{;\mathbf{k}}}$. Also the matrix elements of $\mathbf{r}^e$ are $\mathbf{r}_{{{L}_{\alpha }}{{R}_{\beta }}}^{e}={{\delta }_{\alpha \beta }}{{\xi }_{{{L}_{\alpha }}{{R}_{\beta }}}}$.

\subsubsection{FOOC}
Following a similar approach to the velocity gauge, we can find the optical response in the length gauge. The first-order density matrix in this case is decomposed into two parts, i.e., intraband and interband parts $\varrho^{[1]}=\varrho^{[1]i}+\varrho^{[1]e}$ associated with two components of the position operator defined in Eq.~(\ref{rire}). Assuming that the zeroth-order density matrix is the one in Eq.~(\ref{eqrho}), we can evaluate the first-order density matrix, which includes intraband and interband parts. Using Eq.~(\ref{rhon}) and (\ref{rire}), defining $\mathcal{V}_z(t)=Ee^{-i\omega_1 t}$ and following similar calculations as Ref.~\cite{dabiri2025dynamical}, the matrix elements of the intraband part is
\begin{widetext}
\begin{equation}
\begin{aligned}
  & \varrho _{{{L}_{\alpha }}{{L}_{\beta }}}^{[1]i}=-i{{e}^{-i{{\epsilon }_{\alpha \beta }}t}}E\int_{-\infty }^{t}{{{e}^{i({{\epsilon }_{\alpha \beta }}-{{\omega }_{1}})t'}}{{[{{z}^{i}},{{\varrho }^{[0]}}]}_{{{L}_{\alpha }}{{L}_{\beta }}}}d{{t}^{\prime }}} \\ 
 & =\frac{E{{e}^{-i{{\omega }_{1}}t}}}{{{\omega }_{1}}}{{\delta }_{\alpha \beta }}(i{{\partial }_{{{k}_{z}}}}{{\mathcal{F}}_{\alpha }}+{{\mathcal{F}}_{\alpha }}({{\xi }_{{{L}_{\alpha }}{{R}_{\alpha }}}}-{{\xi }_{{{R}_{\alpha }}{{L}_{\alpha }}}}))  =\frac{iE{{e}^{-i{{\omega }_{1}}t}}}{{{\omega }_{1}}}{{\delta }_{\alpha \beta }}{{\left( {{\mathcal{F}}_{\alpha }} \right)}_{;{{k}_{z}}}},
\end{aligned}
\label{rho1i}
\end{equation}
and the interband part is
\begin{equation}
\begin{aligned}
  & \varrho _{{{L}_{\alpha }}{{L}_{\beta }}}^{[1]e}=-i{{e}^{-i{{\epsilon }_{\alpha \beta }}t}}E\int_{-\infty }^{t}{{{e}^{i({{\epsilon }_{\alpha \beta }}-{{\omega }_{1}})t'}}{{[{{z}^{e}},{{\varrho }^{[0]}}]}_{{{L}_{\alpha }}{{L}_{\beta }}}}d{{t}^{\prime }}} \\ 
 & =-i{{e}^{-i{{\epsilon }_{\alpha \beta }}t}}E\int_{-\infty }^{t}{{{e}^{i({{\epsilon }_{\alpha \beta }}-{{\omega }_{1}})t'}}\times } (z_{{{L}_{\alpha }}{{R}_{\beta }}}^{e}\varrho _{{{L}_{\beta }}{{L}_{\beta }}}^{[0]}-\varrho _{{{L}_{\alpha }}{{L}_{\alpha }}}^{[0]}z_{{{R}_{\alpha }}{{L}_{\beta }}}^{e})d{{t}^{\prime }}=-E\frac{{{e}^{-i{{\omega }_{1}}t}}}{{{\epsilon }_{\alpha \beta }}-{{\omega }_{1}}}({{\mathcal{F}}_{\beta }}z_{{{L}_{\alpha }}{{R}_{\beta }}}^{e}-{{\mathcal{F}}_{\alpha }}z_{{{R}_{\alpha }}{{L}_{\beta }}}^{e}). 
\end{aligned}
\label{rho1e}
\end{equation}
As noted in the main text
  
\begin{equation}
\begin{aligned}
   {{\dsigma }^{[1]}_{xz}}&=\frac{-\text{Tr}({{v}^{x[0]}}{{\varrho }^{[1]}})+\text{Tr}({{v}^{x[0]}}{{\varrho }^{[0]}})\text{Tr}({{\varrho }^{[1]}})}{E{{e}^{-i\omega t}}} =\sum\limits_{\mathbf{k}} {{\dsigma }^{[1]i}}+{{\dsigma }^{[1]e}} 
\label{sig1length}
\end{aligned}
\end{equation}
From Eqs.~(\ref{rho1i}) and (\ref{rho1e}) and using Eq.~(\ref{sig1length}), it is straightforward to obtain the explicit formulas for FOOC in the length gauge
\begin{eqnarray}
  && {{\dsigma }^{[1]i}}=\frac{-1}{{{\omega }_{1}}}\sum\limits_{\alpha }{\left( h_{{{R}_{\alpha }}{{R}_{\alpha }}}^{x}-{{\langle {{h}^{x}}\rangle }_{0}}\mathbf{1}_{{{R}_{\alpha }}{{R}_{\alpha }}} \right)}\nonumber (i{{\partial }_{{{k}_{z}}}}{{\mathcal{F}}_{\alpha }}+{{\mathcal{F}}_{\alpha }}({{\xi }^z_{{{L}_{\alpha }}{{R}_{\alpha }}}}-{{\xi }^z_{{{R}_{\alpha }}{{L}_{\alpha }}}})) \label{sig1i}\\ 
 && {{\dsigma }^{[1]e}}=\sum\limits_{\alpha \beta }{\frac{\left( h_{{{R}_{\beta }}{{R}_{\alpha }}}^{x}-{{\langle {{h}^{x}}\rangle }_{0}}\mathbf{1}_{{{R}_{\beta }}{{R}_{\alpha }}} \right)}{{{\epsilon }_{\alpha \beta }}-{{\omega }_{1}}}} ({{\mathcal{F}}_{\beta }}z_{{{L}_{\alpha }}{{R}_{\beta }}}^{e}-{{\mathcal{F}}_{\alpha }}z_{{{R}_{\alpha }}{{L}_{\beta }}}^{e}) \label{sig1e}.
\end{eqnarray}

\subsubsection{SOOC}
We now evaluate the second-order density matrix. Noting the fact that the position operator splits into intraband and interband parts, the second-order density matrix is composed of four terms $\varrho^{[2]}=\varrho^{[2]ii}+\varrho^{[2]ie}+\varrho^{[2]ei}+\varrho^{[2]ee}$. By subsituting $\mathcal{V}_y=E_2 e^{-i\omega_2 t}$ in Eq.~(\ref{rhon}) and using Eqs.~(\ref{rire}), (\ref{rho1i}) and (\ref{rho1e}), all second-order density matrices can be calculated. The intraband-intraband part is
\begin{equation}
\begin{aligned}
  & \varrho _{{{L}_{\alpha }}{{L}_{\beta }}}^{[2]ii}=-i{{e}^{-i{{\epsilon }_{\alpha \beta }}t}}{{E}_{2}}\int_{-\infty }^{t}{{{e}^{i({{\epsilon }_{\alpha \beta }}-{{\omega }_{2}})t'}}{{[{{y}^{i}},{{\varrho }^{[1]i}}]}_{{{L}_{\alpha }}{{L}_{\beta }}}}d{{t}^{\prime }}} \\ 
 & =-i{{e}^{-i{{\epsilon }_{\alpha \beta }}t}}{{E}_{2}}\int_{-\infty }^{t}{{{e}^{i({{\epsilon }_{\alpha \beta }}-{{\omega }_{2}})t'}}\times } \left[ i{{\partial }_{{{k}_{y}}}}(\varrho _{{{L}_{\alpha }}{{L}_{\beta }}}^{[1]i})+\varrho _{{{L}_{\alpha }}{{L}_{\beta }}}^{[1]i}(\xi _{{{L}_{\alpha }}{{R}_{\alpha }}}^{y}-\xi _{{{R}_{\beta }}{{L}_{\beta }}}^{y}) \right]d{{t}^{\prime }} 
\end{aligned}
\label{}
\end{equation}
All components of the second-order density matrix are derived similarly and can be represented as

\begin{eqnarray}
&&\varrho _{{{L}_{\alpha }}{{L}_{\beta }}}^{[2]ii}=\frac{-E{{E}_{2}}{{e}^{i(-{{\omega }_{1}}-{{\omega }_{2}})t}}}{{{\omega }_{1}}({{\omega }_{1}}+{{\omega }_{2}})}{{\delta }_{\alpha \beta }}{{\left( {{\left( {{\mathcal{F}}_{\alpha }} \right)}_{;{{k}_{z}}}} \right)}_{;{{k}_{y}}}}  \label{rho2ii}\\
 && \varrho _{{{L}_{\alpha }}{{L}_{\beta }}}^{[2]ie}=i{{E}_{2}}E\frac{{{e}^{i(-{{\omega }_{1}}-{{\omega }_{2}})t}}}{{{\epsilon }_{\alpha \beta }}-{{\omega }_{1}}-{{\omega }_{2}}}{{\left( \frac{{{\mathcal{F}}_{\beta }}z_{{{L}_{\alpha }}{{R}_{\beta }}}^{e}-{{\mathcal{F}}_{\alpha }}z_{{{R}_{\alpha }}{{L}_{\beta }}}^{e}}{{{\epsilon }_{\alpha \beta }}-{{\omega }_{1}}} \right)}_{;{{k}_{y}}}}, \label{rho2ie}\\ 
 && \varrho _{{{L}_{\alpha }}{{L}_{\beta }}}^{[2]ei}=\frac{-iE{{E}_{2}}{{e}^{i(-{{\omega }_{1}}-{{\omega }_{2}})t}}\left[ y_{{{L}_{\alpha }}{{R}_{\beta }}}^{e}{{\left( {{\mathcal{F}}_{\beta }} \right)}_{;{{k}_{z}}}}-{{\left( {{\mathcal{F}}_{\alpha }} \right)}_{;{{k}_{z}}}}y_{{{R}_{\alpha }}{{L}_{\beta }}}^{e} \right]}{{{\omega }_{1}}({{\epsilon }_{\alpha \beta }}-{{\omega }_{1}}-{{\omega }_{2}})}, \label{rho2ei}\\ 
 && \varrho _{{{L}_{\alpha }}{{L}_{\beta }}}^{[2]ee}=\sum\limits_{\gamma }{\frac{E{{E}_{2}}{{e}^{i(-{{\omega }_{1}}-{{\omega }_{2}})t}}}{{{\epsilon }_{\alpha \beta }}-{{\omega }_{1}}-{{\omega }_{2}}}\{y_{{{L}_{\alpha }}{{R}_{\gamma }}}^{e}\frac{({{\mathcal{F}}_{\beta }}z_{{{L}_{\gamma }}{{R}_{\beta }}}^{e}-{{\mathcal{F}}_{\gamma }}z_{{{R}_{\gamma }}{{L}_{\beta }}}^{e})}{{{\epsilon }_{\gamma \beta }}-{{\omega }_{1}}}}-\frac{({{\mathcal{F}}_{\gamma }}z_{{{L}_{\alpha }}{{R}_{\gamma }}}^{e}-{{\mathcal{F}}_{\alpha }}z_{{{R}_{\alpha }}{{L}_{\gamma }}}^{e})}{{{\epsilon }_{\alpha \gamma }}-{{\omega }_{1}}}y_{{{R}_{\gamma }}{{L}_{\beta }}}^{e}\}.\label{rho2ee} 
\end{eqnarray}

Thus, SOOC in the length gauge can be derived using the first and second-order density matrices. It can be written as
\begin{equation}
\begin{aligned}
  & {{\dsigma }^{[2]}_{xyz}}=\frac{-\text{Tr}({{v}^{x[0]}}{{\varrho }^{[2]}})}{E{{E}_{2}}{{e}^{-i({{\omega }_{1}}+{{\omega }_{2}})t}}}\text{ }  +\frac{\text{Tr}({{v}^{x[0]}}{{\varrho }^{[0]}})\text{Tr}({{\varrho }^{[2]}})+\text{Tr}({{v}^{x[0]}}{{\varrho }^{[1]}})\text{Tr}({{\varrho }^{[1]}})}{E{{E}_{2}}{{e}^{-i({{\omega }_{1}}+{{\omega }_{2}})t}}} \\ 
 & =\frac{\sum\limits_{\alpha \beta \mathbf{k}}{(-h_{{{R}_{\beta }}{{R}_{\alpha }}}^{x}+{{\langle {{h}^{x}}\rangle }_{0}}\mathbf{1}_{{{R}_{\beta }}{{R}_{\alpha }}})\varrho _{{{L}_{\alpha }}{{L}_{\beta }}}^{[2]}}}{E{{E}_{2}}{{e}^{-i({{\omega }_{1}}+{{\omega }_{2}})t}}}\text{ } \text{+}\sum\limits_{\alpha \beta \mathbf{k}}{h_{{{R}_{\beta }}{{R}_{\alpha }}}^{x}\varrho _{{{L}_{\alpha }}{{L}_{\beta }}}^{[1]}{{|}_{({{\omega }_{1}},z)\to ({{\omega }_{2}},y)}}\frac{\text{Tr}({{\varrho }^{[1]}})}{E{{e}^{-i{{\omega }_{1}}t}}}}. \\ 
\label{sig2length}
\end{aligned}
\end{equation}
\end{widetext}
 \section{Gauge invariance}
The gauge invariance of the OC formulas  (i.e., the invariance with respect to change of the phases of Bloch wave functions) presented in this paper (Eqs.~(\ref{sig1n}) and (\ref{sigmaha})) can be proved easily. Note that all the quantities used in OC formulas are at the same  $\mathbf{k}$ point. If one multiplies the eigenvector $|R_\alpha \rangle$ with a phase factor, $|R_\alpha \rangle \rightarrow |R_\alpha \rangle e^{i\varphi_\alpha}$, then the complex conjugate phase factor should be multiplied to the left eigenvector $\langle L_\alpha | \rightarrow \langle L_\alpha | e^{-i\varphi_\alpha}$ in order to keep the biorthonormalization intact. For example, after a gauge transformation, according to Eq.~(\ref{sig1n}), the OC $\sigma^{[1]}_{\{010\}}$ would become
\begin{equation}
\begin{aligned}
  & \sigma _{\{010\}}^{[1]}=-\frac{i}{{{\omega }_{1}}}\sum\limits_{\alpha \beta }{\frac{{{e}^{i({{\varphi }_{\alpha }}-{{\varphi }_{\beta }})}}h_{{{R}_{\beta }}{{R}_{\alpha }}}^{x}({{\mathcal{F}}_{\beta }}{{e}^{i({{\varphi }_{\beta }}-{{\varphi }_{\alpha }})}}h_{{{L}_{\alpha }}{{R}_{\beta }}}^{z})}{{{\epsilon }_{\alpha \beta }}-{{\omega }_{1}}}}, \\ 
 & +\frac{i}{{{\omega }_{1}}}\sum\limits_{\alpha \beta }{\frac{{{e}^{i({{\varphi }_{\alpha }}-{{\varphi }_{\beta }})}}h_{{{R}_{\beta }}{{R}_{\alpha }}}^{x}({{\mathcal{F}}_{\alpha }}{{e}^{i({{\varphi }_{\beta }}-{{\varphi }_{\alpha }})}}h_{{{R}_{\alpha }}{{L}_{\beta }}}^{z})}{{{\epsilon }_{\alpha \beta }}-{{\omega }_{1}}}}. \\ 
\end{aligned}
\label{}
\end{equation}
The phase factor cancels, and the result is gauge invariant.

Now, let us show the gauge invariance of the covariant derivative defined in Eq.~(\ref{rire}). After adding an arbitrary phase factor and defining $\varphi_{\beta\alpha}\equiv \varphi_{\beta}-\varphi_{\alpha}$ we have
\begin{equation}
\begin{aligned}
  & {{\left( {{e}^{i{{\varphi }_{\beta \alpha }}}}{{\varrho }_{{{L}_{\alpha }}{{L}_{\beta }}}} \right)}_{;\mathbf{k}}}=i(\varphi _{\beta \alpha }^{\prime }){{e}^{i{{\varphi }_{\beta \alpha }}}}{{\varrho }_{{{L}_{\alpha }}{{L}_{\beta }}}}+{{e}^{i{{\varphi }_{\beta \alpha }}}}{{\partial }_{\mathbf{k}}}{{\varrho }_{{{L}_{\alpha }}{{L}_{\beta }}}} \\ 
 & ~~~~~~~~~~~~~~~~~~~~~~+{{\varrho }_{{{L}_{\alpha }}{{L}_{\beta }}}}{{e}^{i{{\varphi }_{\beta \alpha }}}}({{e}^{-i{{\varphi }_{\alpha }}}}\langle {{L}_{\alpha }}|{{\partial }_{\mathbf{k}}}{{e}^{i{{\varphi }_{\alpha }}}}{{R}_{\alpha }}\rangle  \\ 
 &~~~~~~~~~~~~~~~~~~~~~~ -{{e}^{-i{{\varphi }_{\beta }}}}\langle {{R}_{\beta }}|{{\partial }_{\mathbf{k}}}{{e}^{i{{\varphi }_{\beta }}}}{{L}_{\beta }}\rangle ) \\ 
 & =i(\varphi _{\beta \alpha }^{\prime }){{e}^{i({{\varphi }_{\beta }}-{{\varphi }_{\alpha }})}}{{\varrho }_{{{L}_{\alpha }}{{L}_{\beta }}}}+{{e}^{i({{\varphi }_{\beta }}-{{\varphi }_{\alpha }})}}{{\partial }_{\mathbf{k}}}({{\varrho }_{{{L}_{\alpha }}{{L}_{\beta }}}}) \\ 
 &~~~ +{{\varrho }_{{{L}_{\alpha }}{{L}_{\beta }}}}{{e}^{i({{\varphi }_{\beta }}-{{\varphi }_{\alpha }})}}\{i\varphi _{\alpha }^{\prime }-i\varphi _{\beta }^{\prime }+{{\xi }_{{{L}_{\alpha }}{{R}_{\alpha }}}}-{{\xi }_{{{R}_{\beta }}{{L}_{\beta }}}}\} \\ 
 & ={{e}^{i({{\varphi }_{\beta }}-{{\varphi }_{\alpha }})}}{{\left( {{\varrho }_{{{L}_{\alpha }}{{L}_{\beta }}}} \right)}_{;\mathbf{k}}}. \\ 
\end{aligned}
\label{}
\end{equation}
Therefore, the change of phase can be drawn out from the covariant derivative, and it is an easy task to show that the length gauge OC formulas in Eqs.~(\ref{sig1length}) and (\ref{sig20}) are also gauge invariant.

\section{Obtaining length gauge conductivities from velocity gauge conductivities}
Here, we explain how to obtain OC formulas in the length gauge from velocity gauge formulas. The only task here is to explain the matrix elements of high-order derivatives of the Hamiltonian in terms of its first derivative matrix elements. It can be shown for $X,Y\in(L,R)$ that
\begin{equation}
\begin{aligned}
h_{X_\alpha Y_\beta }^{xz}={{\partial }_{z}}h_{X_\alpha Y_\beta }^{x}-i{{[{{\xi }^{z}},{{h}^{x}}]}_{X_\alpha Y_\beta }}.
\end{aligned}
\label{switch1}
\end{equation}
where we have defined $\xi _{X_\alpha X_\gamma^\prime }^{z}\equiv i\langle {{X_\alpha }}|{{\partial }_{{{k}_{z}}}}{{X^\prime_\gamma }}\rangle $ as a generalize Berry connection. The proof of the above equation is derived easily, and we write it for the case $X=Y=L$:
\begin{equation}
\begin{aligned}
 &  h_{L_\alpha L_\beta }^{xz}=\langle { {L_\alpha }} |{{\partial }_{{{k}_{z}}}}{{\partial }_{{{k}_{x}}}}H|{ L_{\beta }} \rangle ={{\partial }_{{{k}_{z}}}}\langle { L_{\alpha }} |{{\partial }_{{{k}_{x}}}}H|{ L_{\beta }} \rangle -I \\ 
& I= \langle {{\partial }_{{{k}_{z}}}}{L_{\alpha }} |{{\partial }_{{{k}_{x}}}}H|{L_{\beta }} \rangle +\langle {L_{\alpha }} |{{\partial }_{{{k}_{x}}}}H|{{\partial }_{{{k}_{z}}}}{L_{\beta }} \rangle  \\ 
& = -i\sum\limits_{\gamma }{\left( -\xi _{L_\alpha R_\gamma }^{z}h_{L_\gamma L_\beta }^{x}+h_{L_\alpha L_\gamma }^{x}\xi _{R_\gamma L_\beta }^{z} \right)}=i{{[{{\xi }^{z}},{{h}^{x}}]}_{L_\alpha L_\beta }} 
\end{aligned}
\label{switch2}
\end{equation}
where for obtaining the third line we inserted the complete set $\mathbf{1}=\sum\nolimits_{\gamma }{|{R_{\gamma }} \rangle \langle {L_{\gamma }} |}$ and also used the fact that $\langle {{\partial }_{{{k}_z}}}{L_{\alpha }} |{R_{\gamma }} \rangle +\langle {L_{\alpha }} |{{\partial }_{{{k}_z}}}{R_{\gamma }} \rangle ={{\partial }_{{{k}_z}}}{{\delta }_{\alpha \gamma }}=0$.

A useful new sum-rule can be obtained by substituting $h^x\rightarrow \mathbf{1}$ in Eq.~(\ref{switch1}) as follows
\begin{equation}
\begin{aligned}
0={{\partial }_{z}}\mathbf{1}_{X_\alpha Y_\beta } -i{{[{{\xi }^{z}},{\mathbf{1}}]}_{X_\alpha Y_\beta }}.
\end{aligned}
\label{switch3}
\end{equation}

By substituting the relation (\ref{switch1}) into the first equation of (\ref{sig1n}), performing integration by parts, and exploiting the sum-rule (\ref{switch3}), we obtain the FOOC formula in the length gauge.

To determine the second-order response in the length gauge, we need to express the second-order and third-order derivatives of the Hamiltonian (as in Eq.~(\ref{switch2})) in terms of its first derivative in the formula (\ref{sigmaha}). The matrix elements of the third-order derivative of the Hamiltonian can be related to the second-order derivative in the following manner:

\begin{equation}
\begin{aligned}
h_{\alpha \beta }^{xyz}={{\partial }_{z}}h_{\alpha \beta }^{xy}-i{{[{{\xi }^{z}},{{h}^{xy}}]}_{\alpha \beta }}.
\end{aligned}
\label{switch4}
\end{equation}
 
Let us explicitly show how to obtain the linear response formula for the length gauge Eq.~(\ref{sig1length}) from the velocity gauge formula Eq.~(\ref{sig1n}). Using Eq.~(\ref{switch1}) one can rewrite the first term in OC of (\ref{sig1n}) as
\begin{widetext}
\begin{equation}
\begin{aligned}
  & \sum\limits_{\mathbf{k}}{\,}\sigma _{\{100\}}^{[1]}=\sum\limits_{\alpha \mathbf{k}}{\,}\frac{i}{{{\omega }_{1}}}{{\mathcal{F}}_{\alpha }}h_{{{R}_{\alpha }}{{R}_{\alpha }}}^{xz}=\sum\limits_{\alpha \mathbf{k}}{\,}\frac{-ih_{{{R}_{\alpha }}{{R}_{\alpha }}}^{x}{{\partial }_{{{k}_{\text{z}}}}}{{\mathcal{F}}_{\alpha }}}{{{\omega }_{1}}}+\frac{{{\mathcal{F}}_{\alpha }}{{[{{\xi }^{z}},{{h}^{x}}]}_{{{R}_{\alpha }}{{R}_{\alpha }}}}}{{{\omega }_{1}}} \\ 
 & =\sum\limits_{\alpha \mathbf{k}}{\,}\frac{h_{{{R}_{\alpha }}{{R}_{\alpha }}}^{x}(-i{{\partial }_{{{k}_{\text{z}}}}}{{\mathcal{F}}_{\alpha }}+{{\mathcal{F}}_{\alpha }}\xi _{{{R}_{\alpha }}{{L}_{\alpha }}}^{z}-\xi _{{{L}_{\alpha }}{{R}_{\alpha }}}^{z})}{{{\omega }_{1}}}+X, \\ 
 & X=\sum\limits_{\alpha \beta \mathbf{k}}{\,}\frac{{{\mathcal{F}}_{\alpha }}(z_{{{R}_{\alpha }}{{L}_{\beta }}}^{e}h_{{{R}_{\beta }}{{R}_{\alpha }}}^{x}-h_{{{R}_{\alpha }}{{R}_{\beta }}}^{x}z_{{{L}_{\beta }}{{R}_{\alpha }}}^{e})}{{{\omega }_{1}}}=\sum\limits_{\alpha \beta \mathbf{k}}{\,}\frac{h_{{{R}_{\beta }}{{R}_{\alpha }}}^{x}({{\mathcal{F}}_{\alpha }}h_{{{R}_{\alpha }}{{L}_{\beta }}}^{z}-{{\mathcal{F}}_{\beta }}z_{{{L}_{\alpha }}{{R}_{\beta }}}^{e})}{i{{\epsilon }_{\alpha \beta }}{{\omega }_{1}}} \\ 
\end{aligned}
\label{expand1}
\end{equation}
where in the first line we took integration by parts using the fact that the Brillouin zone is a closed path. The third line above cancels the divergence in $\sigma _{\{010\}}^{[1]}$ defined in Eq.~(\ref{sig1n}). Using the identity (\ref{switch3}) one can write 
\begin{equation}
\begin{aligned}
  & \text{ }\sum\limits_{\mathbf{k}}{}\sigma _{\{100\}}^{[1]}=\sum\limits_{\alpha \mathbf{k}}{}\frac{i}{{{\omega }_{1}}}{{\mathcal{F}}_{\alpha }}h_{{{R}_{\alpha }}{{R}_{\alpha }}}^{xz}-\sum\limits_{\alpha \mathbf{k}}{}\frac{i}{{{\omega }_{1}}}{{\mathcal{F}}_{\alpha }}{{\langle {{h}^{x}}\rangle }_{0}}\mathbf{0}_{{{R}_{\alpha }}{{R}_{\alpha }}}^{xz} \\ 
 & \sum\limits_{\alpha \mathbf{k}}{}\frac{i{{\langle {{h}^{x}}\rangle }_{0}}}{{{\omega }_{1}}}{{\mathcal{F}}_{\alpha }}\mathbf{0}_{{{R}_{\alpha }}{{R}_{\alpha }}}^{xz}=\sum\limits_{\alpha \mathbf{k}}{}\frac{-i{{\langle {{h}^{x}}\rangle }_{0}}{{\mathcal{F}}_{\alpha }}{{\partial }_{{{k}_{\text{z}}}}}\mathbf{1}_{{{R}_{\alpha }}{{R}_{\alpha }}}^{}}{{{\omega }_{1}}}+\frac{{{\langle {{h}^{x}}\rangle }_{0}}{{\mathcal{F}}_{\alpha }}{{[{{\xi }^{z}},\mathbf{1}]}_{{{R}_{\alpha }}{{R}_{\alpha }}}}}{{{\omega }_{1}}}\text{ } \\ 
 & =\sum\limits_{\alpha \mathbf{k}}{}\frac{-i{{\langle {{h}^{x}}\rangle }_{0}}\mathbf{1}_{{{R}_{\alpha }}{{R}_{\alpha }}}^{}{{\partial }_{{{k}_{\text{z}}}}}{{\mathcal{F}}_{\alpha }}}{{{\omega }_{1}}}+\frac{{{\langle {{h}^{x}}\rangle }_{0}}{{\mathcal{F}}_{\alpha }}{{[{{\xi }^{z}},\mathbf{1}]}_{{{R}_{\alpha }}{{R}_{\alpha }}}}}{{{\omega }_{1}}} \\ 
 & =\sum\limits_{\alpha \mathbf{k}}{}\frac{{{\langle {{h}^{x}}\rangle }_{0}}\mathbf{1}_{{{R}_{\alpha }}{{R}_{\alpha }}}^{x}(-i{{\partial }_{{{k}_{\text{z}}}}}{{\mathcal{F}}_{\alpha }}+{{\mathcal{F}}_{\alpha }}\xi _{{{R}_{\alpha }}{{L}_{\alpha }}}^{z}-\xi _{{{L}_{\alpha }}{{R}_{\alpha }}}^{z})}{{{\omega }_{1}}}+Y\text{ } \\ 
 & Y=\sum\limits_{\alpha \beta \mathbf{k}}{}\frac{{{\langle {{h}^{x}}\rangle }_{0}}{{\mathcal{F}}_{\alpha }}(z_{{{R}_{\alpha }}{{L}_{\beta }}}^{e}\mathbf{1}_{{{R}_{\beta }}{{R}_{\alpha }}}^{}-\mathbf{1}_{{{R}_{\alpha }}{{R}_{\beta }}}^{}z_{{{L}_{\beta }}{{R}_{\alpha }}}^{e})}{{{\omega }_{1}}}=\sum\limits_{\alpha \beta \mathbf{k}}{}\frac{{{\langle {{h}^{x}}\rangle }_{0}}\mathbf{1}_{{{R}_{\beta }}{{R}_{\alpha }}}^{}({{\mathcal{F}}_{\alpha }}h_{{{R}_{\alpha }}{{L}_{\beta }}}^{z}-{{\mathcal{F}}_{\beta }}z_{{{L}_{\alpha }}{{R}_{\beta }}}^{e})}{i{{\epsilon }_{\alpha \beta }}{{\omega }_{1}}} 
\end{aligned}
\label{expand2}
\end{equation}
where in the third line we took integration by parts and used the following fact  
$$\sum\limits_{\alpha \mathbf{k}}{\,}{{\mathcal{F}}_{\alpha }}\mathbf{1}_{{{R}_{\alpha }}{{R}_{\alpha }}}^{\,}{{\partial }_{{{k}_{\text{z}}}}}{{\langle {{h}^{x}}\rangle }_{0}}=\sum\limits_{\mathbf{k}}{\,}\text{Tr}({{\rho }^{[0]}}){{\partial }_{{{k}_{\text{z}}}}}{{\langle {{h}^{x}}\rangle }_{0}}=\sum\limits_{\mathbf{k}}{\,}{{\partial }_{{{k}_{\text{z}}}}}{{\langle {{h}^{x}}\rangle }_{0}}=0 $$ 
which originates from the Brillouin zone being a closed path. Substituting Eqs.~(\ref{expand1}) and (\ref{expand2}) in Eq.~(\ref{sig1n}) one can obtain Eq.~(\ref{sig1length}).

\end{widetext}

\section{Non-Hermitian Su-Schrieffer-Heeger model}
The SSH model is the simplest one-dimensional tight-binding model exhibiting topological phases, originally proposed for electronic states in polyacetylene \cite{sshref}. It is defined on a bipartite lattice with nearest-neighbor intracell and intercell hoppings, $t$ and $\Delta$, respectively, with $t, \Delta \in \mathbb{R}$. In a non-Hermitian extension of this model, we introduce an imaginary onsite potential, $i\gamma$, where $\gamma \in \mathbb{R}$ represents balanced gain and loss. The momentum space representation of the Hamiltonian is 
\begin{equation}
\begin{aligned}
\mathcal{H}_{SSH}=t {\sigma }_{x}+{\Delta}(\cos k {\sigma }_{x}-\sin k {\sigma }_{y})+i\gamma \sigma_z.
\end{aligned}
\label{hssh}
\end{equation}

\begin{figure}
\includegraphics[width=\linewidth,trim={0 0 0.25cm 0}, clip]{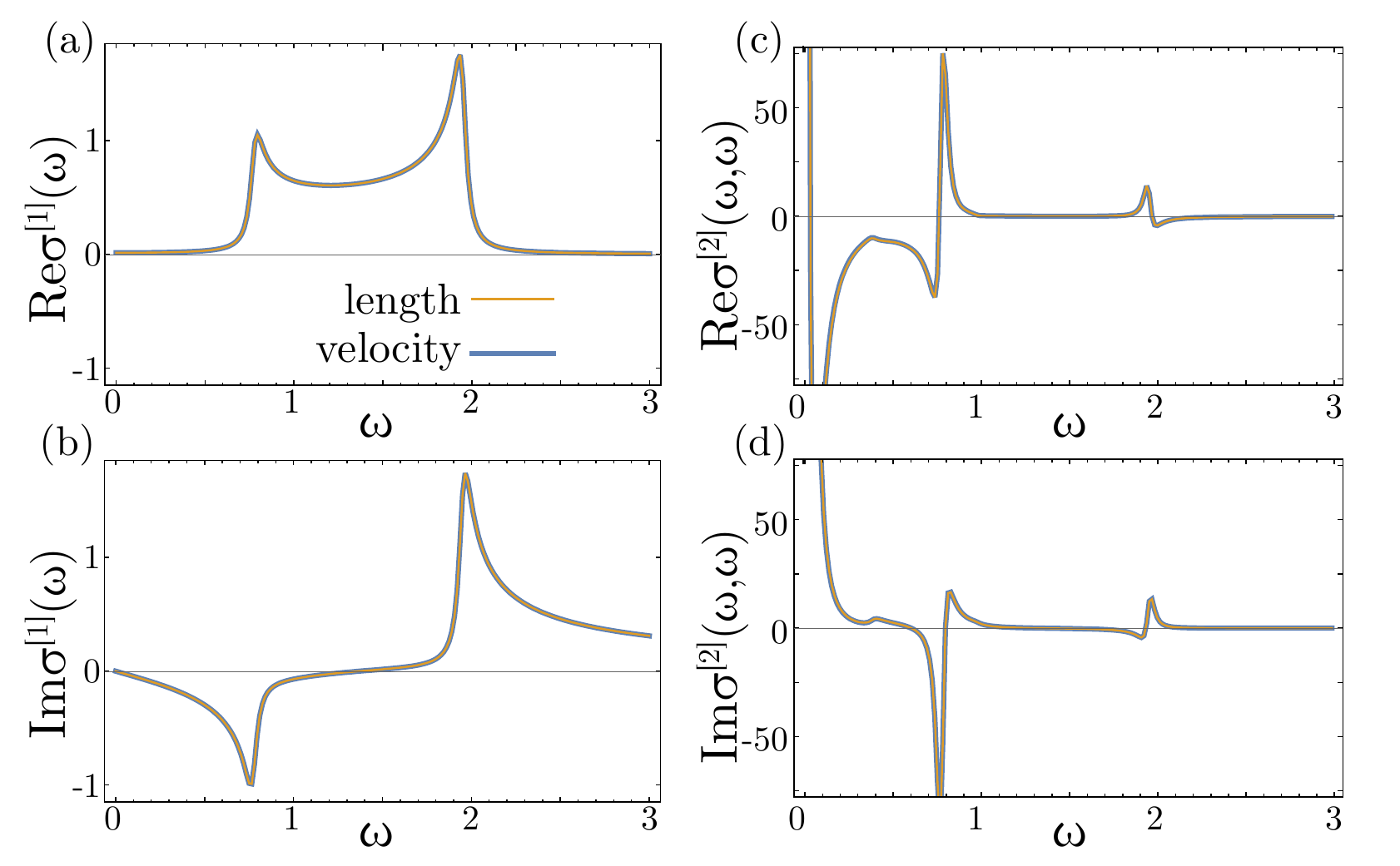} 
\caption{(Color online)  Panels (a) and (b) display the linear response: both real and imaginary parts show well-defined resonances at interband transition frequencies, with perfect agreement between gauges, confirming the gauge consistency of the formalism for non-Hermitian qubit Hamiltonian (\ref{hqubit}) with parameters $t=0.2, \gamma=0.7$. No low-frequency divergence occurs, as expected for a gapped, fully filled band structure.
Panels (c) and (d) reveal the second-order response. Compared to the linear case, the number of resonant features doubles, corresponding to both one-photon and two-photon processes.  Singularities appear, stemming from interband–interband contributions that have no Hermitian analog. Notably, a low-frequency divergence emerges in the SOOC due to these terms, which reflects the absence of conventional cancellation mechanisms in pseudo-Hermitian systems. This divergence is a direct illustration of the modified sum rules and could be probed in waveguide experiments via low-frequency harmonic generation. The excellent agreement between length and velocity gauge results across all frequencies demonstrates the robustness of the gauge-invariant formalism developed here.}
\label{sshfig}
\end{figure}
The energy dispersion of Eq.~(\ref{sshfig}) is $\epsilon_{SSH}=\pm \sqrt{t^2+\Delta^2+2t \Delta \cos{k}-\gamma^2}$, therefore for $|\gamma|<|t\pm \Delta|$ the system is in the $n \mathcal{PT}$ phase with a completely real spectrum. The non-Hermitian SSH model has been realized experimentally in photonic waveguides \cite{slootman2024breaking}, where non-Hermiticity is engineered via controlled gain and loss in waveguide arrays. It has also been implemented in mechanical systems \cite{li2024observation} and electrical circuits \cite{liu2021non}. Using our formalism, the FOOC can be calculated in both the length and velocity gauges via Eqs.~(\ref{sig1length}) and (\ref{sig1n}) and the second-order response in the length and velocity gauge by Eqs.~(\ref{sig20}) and (\ref{sigmaha}), respectively. Due to the complexity of the integrals involved, we perform numerical evaluations and report analytical results only for the values of selected parameters. The linear response is shown in Fig.~\ref{sshfig}(a),(b). No divergence appears at low frequencies, which is justified as the system is gapped and bands are fully filled. The real part of OC remains positive, and the derivative of the real (imaginary) part at resonant frequencies is minimum (maximum), similar to Hermitian systems. As shown in this figure, the linear responses in both gauges are fully consistent. Analytical results for special parameters $\gamma=t=1/2, \omega=1$ can be derived as 
\begin{equation}
   {{\sigma }^{[1]}}=\frac{i[14\sqrt{2}-16\sqrt{7}\text{ArcTanh}(2\sqrt{\frac{2}{7}})]}{56\pi } 
 +\frac{5i\text{ArcTanh}(\frac{2\sqrt{2}}{3})}{8\pi },
\label{}
\end{equation}

The second-order results are shown in Fig.~\ref{sshfig}(c),(d), demonstrating a complete agreement between the two gauges. The number of resonances is twice Fig.~\ref{sshfig}(a), corresponding to one-photon and two-photon resonances. A divergence is seen at low frequencies, which is related to the interband-interband contributions (see Eq.~(\ref{rho2ee})). Our calculations for this model confirm the validity of the formalism presented in this paper.

Despite the linear response, the SOOC of the SSH model shows both one- and two-photon resonances, mainly doubling the number of spectral features. These multi-peak stems from the additional pseudo-Hermitian terms that multiply the first-order response and generate higher-order pole structures. Furthermore, the low-frequency divergence seen in Figs. 2(c),(d) can be traced to interband–interband contributions (Eq. A12), which have no Hermitian counterpart. These features emphasize how gain–loss balance modifies the analytic structure of NL optical response while maintaining gauge consistency.

\end{document}